# Enhanced Heart Sound Classification Using Mel Frequency Cepstral Coefficients and Comparative Analysis of Single vs. Ensemble Classifier Strategies

**Amir Masoud Rahmani[1,+], Amir Haider[2,+], Mohammad Adeli[3], Olfa Mzoughi[4], Entesar Gemeay[5], Mokhtar Mohammadi[6], Hamid Alinejad-Rokny[7,8], Parisa Khoshvaght[9], and Mehdi Hosseinzadeh[9,10,*]**

[1] Future Technology Research Center, National Yunlin University of Science and Technology, Yunlin, Taiwan
[2] Department of Computer Science and Engineering, Sejong University, Seoul 05006, Republic of Korea
[3] Department of Biomedical Engineering, Dezful Branch, Islamic Azad University, Dezful, Iran
[4] Department of Computer Sciences, College of Computer Engineering & Sciences, Prince Sattam bin Abdulaziz University, Al-Kharj 11942, Saudi Arabia
[5] Department of Computer Engineering, Computer and Information Technology College, Taif University, Taif, Saudi Arabia
[6] Department of Information Technology, College of Engineering and Computer Science, Lebanese French University, Kurdistan Region, Iraq
[7] UNSW BioMedical Machine Learning Lab (BML), The Graduate School of Biomedical Engineering, UNSW Sydney, 2052, NSW, Australia\
[8] Tyree Institute of Health Engineering (IHealthE), UNSW Sydney, 2052, NSW, Australia
[9] Institute of Research and Development, Duy Tan University, Da Nang, Vietnam
[10] School of Medicine and Pharmacy, Duy Tan University, Da Nang, Vietnam

[+]These authors contributed equally to this work.

*Corresponding Authors: Mehdi Hosseinzadeh mehdihosseinzadeh@duytan.edu.vn

**Abstract**
This paper explores the efficacy of Mel Frequency Cepstral Coefficients (MFCCs) in detecting abnormal heart sounds using two classification strategies: a single classifier and an ensemble classifier approach. Heart sounds were first pre-processed to remove noise and then segmented into S1, systole, S2, and diastole intervals, with thirteen MFCCs estimated from each segment, yielding 52 MFCCs per beat. Finally, MFCCs were used for heart sound classification. For that purpose, in the single classifier strategy, the MFCCs from nine consecutive beats were averaged to classify heart sounds by a single classifier (either a support vector machine (SVM), the k nearest neighbors (kNN), or a decision tree (DT)). Conversely, the ensemble classifier strategy employed nine classifiers (either nine SVMs, nine kNN classifiers, or nine DTs) to individually assess beats as normal or abnormal, with the overall classification based on the majority vote. Both methods were tested on a publicly available phonocardiogram database. The heart sound classification accuracy was 91.95% for the SVM, 91.9% for the kNN, and 87.33% for the DT in the single classifier strategy. Also, the accuracy was 93.59% for the SVM, 91.84% for the kNN, and 92.22% for the DT in the ensemble classifier strategy. Overall, the results demonstrated that the ensemble classifier strategy improved the accuracies of the DT and the SVM by 4.89% and 1.64%, establishing MFCCs as more effective than other features, including time, time-frequency, and statistical features, evaluated in similar studies.

# 1. Introduction

The mechanical activities of the heart and blood flow generate heart sounds [1]. The graphical representation of the heart sounds is usually called a phonocardiogram (PCG). A phonocardiogram typically comprises four components: S1 sound, systole, S2 sound, and diastole. However, there can be other sounds in a PCG [1].

Phonocardiograms can be used to develop assistive intelligent systems to detect cardiovascular diseases [2-5]. In general, such a system is composed of four steps: preprocessing, segmentation, feature extraction, and classification [5, 6]. These steps are briefly reviewed below, but an exhaustive review of these techniques can be found here [5, 7].

Preprocessing usually involves removing undesirable noises, artifacts, and spikes. Some of the techniques that have been used in other studies for PCG preprocessing include the normalization of PCGs to have zero mean [8], low-pass filtering [9, 10], high-pass filtering from 10 Hz and normalization [11], band-pass filtering from 40 Hz to 400 Hz [12], band-pass filtering from 20 Hz to 400 Hz [12, 13], band-pass filtering from 5 Hz to 700 Hz [14], band-pass filtering from 2 Hz to 100 Hz, and discrete wavelet transform (DWT) [15]. Sometimes, no preprocessing is applied [16].

Segmentation aims to find the S1, systole, S2, and diastole intervals in a PCG signal. Schmidt et al. (2010) used a duration-dependent hidden Markov model (DHMM) for PCG segmentation [17]. This method was extended by Springer et al. (2015) [18] using hidden semi-Markov models (HSMM) and logistic regression. This method has been adopted in many studies [3, 12, 13]. In [14], Mel-Scaled Wavelet Transform (MSWT) and dynamic thresholding were used for PCG segmentation while durations of systole and diastole were analyzed to find S1 and S2 in [10]. The PCG segmentation [11] was based on Shannon energy, envelope smoothing, and peak finding. Similarly, Jaros et al. (2023) [15] applied $3^{rd}$-order Shannon energy, envelope detection by low-pass filtering, thresholding, and the k-means algorithm. The PCG segmentation method proposed by Alonso-Arévalo et al. (2021) was based on spectral change detection and genetic algorithms [19]. Many of the PCG segmentation methods are reviewed here [20]. It is also necessary to mention that no segmentation strategies were used in some previous studies [8, 9, 16].

Quantitative features are extracted from the PCG segments in the third step of PCG processing. Various types of PCG features have been used in different applications. These types include time-domain features [3, 10-14], spectral features [8, 11-13], time-frequency features [3, 10-12, 14], time-scale features [8, 11, 13], Mel Frequency Cepstral Coefficients [3, 10, 11, 14, 21], and features estimated by convolutional neural networks (CNNs) [9]. Some feature selection strategies, such as linear discriminant analysis [10], correlation-based feature selection [13], and genetic algorithms, [11] have been used for dimensionality reduction.

The fourth step of PCG processing involves training and testing a classification model to detect underlying diseases [5, 6]. Classification methods that have been used in applications of PCGs include convolutional neural networks (CNNs) [16, 22], artificial neural networks (ANNs) [3, 9, 10, 14], deep neural networks [12], the k nearest neighbors algorithm (kNN) [3, 11, 14], decision trees (DT) [3, 8], long short-term memory (LSTM) networks [3], ensemble classifiers [3, 13], support vector machines (SVMs)[14], and hidden Markov models (HMMs) [23].

This study aimed to investigate the performance of MFCCs in discriminating abnormal PCGs from normal ones. MFCCs were first used in speech processing applications [24], but later in other applications such as PCG processing. MFCCs were selected because they are weakly-correlated and highly discriminating features of audio signals, providing compact spectral representations successfully used in speech processing applications [25]. Despite that, they have performed modestly in some PCG processing applications [3, 10]. Therefore, two classification strategies, referred to as the single-classifier and the ensemble-classifier strategies, are presented to enhance the performance of MFCCs for PCG classification. In the single-classifier strategy, the MFCCs extracted from different PCG beats are first averaged, and the mean MFCCs are then used to classify PCGs. However, in the ensemble-classifier strategy, MFCCs are used by an ensemble of 9 classifiers to classify PCG beats into normal/abnormal beats. In the end, if most beats are classified as normal, the PCG is considered normal; otherwise, it is abnormal. Both strategies were tested on a publicly available database of PCG signals. The results showed that the ensemble-classifier strategy could classify PCGs with a higher accuracy.

The rest of this article is organized as follows: Section 2 describes the PCG database, MFCC estimation, and the two classification strategies. Section 3 presents the results in detail. Section 4 discusses the results and compares them with similar studies. Section 5 presents this study's conclusions.

## 2. Materials and methods
### 2.1. Heart sounds database
The PhysioNET CinC 2016 PCG database [26, 27] was used to evaluate the PCG classification methods proposed in this paper. The signals of this database have been collected from healthy subjects and patients with such heart diseases as heart valve defects and coronary artery problems. The signals lasted from 5 to 120 seconds and were recorded at a rate of 2000 samples/s [26]. The database contains 6 datasets (labeled A to F) that contain 3153 signals (2488 from the healthy subjects and 665 from the patients). Dataset A (including 490 signals) was used to train the segmentation model introduced in Datasets B to F, which were used to evaluate our single-classifier and ensemble-classifier strategies described in section 2.2.4. There are a total of 2744 signals in datasets B to F. Two hundred sixty-two of them that are labeled as "uncertain" are noisy signals and, therefore, were ignored. There remained 2482 signals (including 296 signals from the patient class and 2186 from the healthy class). Only signals with at least 9 PCG beats were used in this study. Among the remaining 2482 signals, 2137 PCGs (including 218 abnormal and 1919 normal PCGs) met this criterion and, therefore, were used to evaluate our single-classifier and ensemble-classifier strategies as described in section 2.2.4.

### 2.2 The proposed method for the classification of heart sounds
The proposed method for heart sound classification includes four stages: pre-processing, segmentation, feature extraction, and classification. These stages are explained in detail as follows.

#### 2.2.1. Preprocessing
All the signals are first resampled from 2000 Hz to 1000 Hz to reduce the computational cost. They are then preprocessed using a band-pass filter from 25 to 400 Hz to remove low-frequency and high-frequency noises. Similar settings were used in other studies [10, 12, 28].

#### 2.2.2. Segmentation
The segmentation aims to find the time intervals of the S1 and S2 sounds and the intervals of the systole and diastole regions in a PCG signal. Springer et al. (2015) [18] proposed a sophisticated supervised method for PCG segmentation. This method includes a feature extraction step followed by a four-class classifier (S1, systole, S2, and diastole). The features were extracted using the Hilbert envelope, power spectrum density, and wavelet transform. The classifier is based on the HSMM and logistic regression [18]. The classifier assigns labels 1, 2, 3, and 4 to the PCG samples that belong to S1, systole, S2, and diastole, respectively.

### 2.2.3. Feature extraction
In this step, MFCCs are estimated for PCG signals. The MFCCs represent the short-term power spectrum of an audio signal. The Mel scale is used to approximate the performance of the auditory system, which uses a non-linear frequency scale instead of a linear one. Estimation of MFCCs involves the following steps [29]:
1) The PCG signal is divided into frames.
2) Discrete Fourier Transform (DFT) of each PCG frame is computed.
3) The power spectrum of each frame is estimated using a Mel filter bank.
4) The logarithm of the power coefficients is calculated.
5) The discrete cosine transformation of the log power coefficients is computed.

To extract the MFCCs, we first divide the PCG signal into 24 ms frames. Adjacent frames overlap by 18 ms. Next, each frame is multiplied by a Hamming window, and then the 64-point DFT coefficients of the windowed frame are computed. Assuming that $X_i[k]$ denotes the DFT coefficients of the $i^{th}$ windowed frame of the PCG signal, we estimate the power spectrum of this frame as $|X_i[k]|^2/N$. Then we calculate the power of this frame within all Mel bands. For that purpose, we use M=20 triangular filters [10]. The frequency response of the $m^{th}$ filter, $H_m(k)$, is defined as follows [29]:

$$H_m(k) = \begin{cases} 0, & k < k_f(m) \\ \dfrac{k - k_f(m)}{k_f(m+1) - k_f(m)}, & k_f(m) \le k < k_f(m+1) \\ \dfrac{k_f(m+2) - k}{k_f(m+2) - k_f(m+1)}, & k_f(m+1) \le k \le k_f(m+2) \\ 0. & k > k_f(m+2) \end{cases}, \quad 0 \le m \le M-1 \qquad (1)$$

Where the variable $k_f(m)$ is the index for the center frequency of the $m^{th}$ triangular filter. A total of cap M plus 2 frequencies are required to design M filters. The relationship between the frequency in the Mel scale ($f_{mel}$) and the frequency in Hertz is calculated from the following equation [29]:

$$f_{mel} = 2595 \log_{10}(1 + \frac{f}{700}) \qquad (2)$$

The minimum and maximum Mel frequencies are calculated for $f_{min} = 0$ and $f_{max} = 400$ Hz using equations (2). Afterwards, we find $M + 2$ equally-spaced Mel frequencies from the minimum to the maximum Mel frequencies. The obtained $M + 2$ Mel frequencies are converted back to the Hz scale using the inverse of equation (2). With the $M + 2$ frequencies required for designing $M$ triangular filters available, the $k_f$ index for the $j^{th}$ frequency is calculated as :

$$k_f(j) = \left[\frac{f(j)}{f_s} N\right], \quad 0 \le j \le M + 1 \qquad (3)$$

Where $f(j)$ is one of the $M + 2$ designed frequencies in Hz, $f_s$ is the sampling frequency, and $N$ is the number of DFT coefficients. Now, the frequency response of these filters can be obtained using equation (1).

The power of the $i^{th}$ frame in the $m^{th}$ Mel band, $P_i[m]$, is estimated as:

$$P_i[m] = 10 \log_{10}\left(\frac{1}{N} \sum_{k=0}^{\frac{N}{2}} |X_i[k]|^2 |H_m[k]|\right), \quad 0 \leq m \leq M - 1 \qquad (4)$$

Finally, the MFCCs $C_i[k']$ of the $i^{th}$ frame are computed from $P_i[m]$ by the type II discrete cosine transformation (DCT-II) [29]:

$$C_i[k'] = \sum_{m=0}^{M-1} P_i[m] \cos\left(\frac{\pi}{M}\left(m + \frac{1}{2}\right)k'\right), \quad 0 \leq k' \leq M - 1 \qquad (5)$$

The number of MFCCs is usually between 12 and 20 [3, 11]. In a nutshell, for a 24-ms frame of the PCG signal, 20 MFCCs are calculated, but only the first 13 coefficients were used in this research. As shown in Figure 1, by averaging the MFCCs obtained for all the frames belonging to the S1 sound, 13 features are obtained. Similarly, 13 features are calculated for each of the other segments, i.e. systole, S2, and diastole. Therefore, 52 MFCC features are obtained for a given PCG beat.

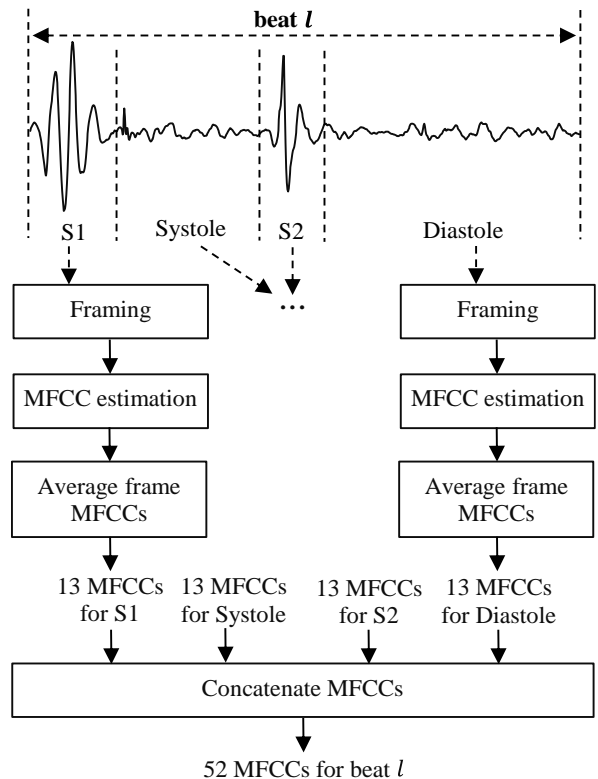

Figure 1. Feature extraction: 13 MFFCs are extracted from each interval (i.e., S1, systole, S2, and diastole) of a given PCG beat, summing up to 52 MFCCs for that beat.

### 2.2.4 Classification
This research used two classification strategies to discriminate normal (healthy) PCGs from abnormal (pathological) ones. These two strategies are referred to as the single-classifier and the ensemble-classifier. In the single-classifier strategy (Figure 2.A), the MFCCs for the first 9 PCG beats are first averaged and then fed to a classifier. Since there are 52 features per beat, averaging results in 52 mean MFCCs (Figure 2.A), based on which the classifier determines whether the PCG signal is normal or abnormal. Three classifier types were used in this strategy: $k$ nearest neighbors (kNN), support vector machine (SVM), and decision tree (DT).

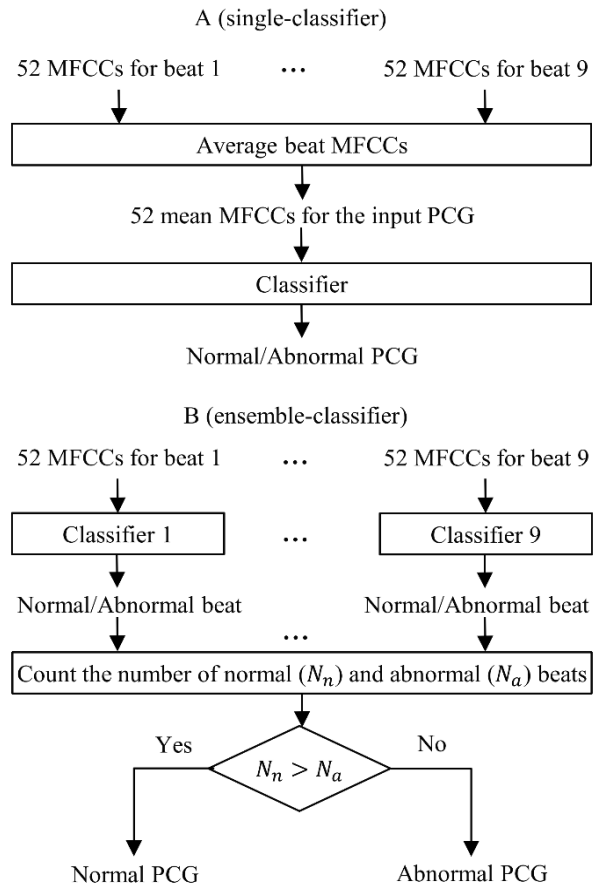

Figure 2. Classification strategies. A: Single-classifier strategy: the MFCCs extracted from the first 9 beats of a PCG are averaged and fed to a single classifier, B: Ensemble-classifier strategy: nine classifiers are used separately to distinguish normal beats from abnormal ones. In the end, if the number of the normal beats is more than the abnormal beats, the PCG signal is decided to be normal; otherwise, it is abnormal.

In the ensemble-classifier strategy (Figure 2.B), the 52 MFCCs of a given beat are fed to a distinct classifier. Since only the first 9 beats of a PCG signal are used, there are 9 different classifiers. Each classifier decides whether its respective input beat is normal or abnormal. In the end, if more normal beats are predicted by the 9 classifiers altogether, the PCG is decided to be normal, otherwise abnormal (Figure 2.B). All nine classifiers are the same type (i.e., kNN, SVM, or DT). For the kNN, k was considered 1, 3, 5, and 7. Also, linear, Gaussian, and polynomial kernels were considered for the support vector machine.

As explained in section 2.1, we selected 218 abnormal and 1919 normal PCG signals with at least 9 beats to evaluate the proposed methods. The first 9 beats were used for MFCC estimation. To balance the dataset, 218 normal PCGs were randomly selected and used alongside the 218 abnormal ones to train and test the classifiers.

Ten-fold cross-validation was used to evaluate both classification strategies. During each fold, we calculated four parameters of accuracy ($Acc$), sensitivity ($Se$), specificity ($Sp$), and modified accuracy ($MAcc$) by:

$$Acc = \frac{TP + TN}{TP + FN + FP + TN} \quad (6)$$

$$Se = \frac{TP}{TP + FN} \quad (7)$$

$$Sp = \frac{TN}{FP + TN} \quad (8)$$

$$MAcc = \frac{Se + Sp}{2} \quad (9)$$

Where $TP$ is the number of patients who were correctly classified as patients, $FN$ is the number of patients who were wrongly classified as healthy, $FP$ is the number of healthy subjects who were wrongly classified as patients, and $TN$ is the number of healthy people who were correctly classified as healthy subjects. The average parameters across the 10 folds were calculated in the end. The 10-fold cross-validation was repeated 50 times. Each time, a random set of 218 normal signals was selected and concatenated with the 218 abnormal PCGs. The average results across the 50 runs are reported in section 3.

It should be noted that all analyses, including feature extraction and classification strategies, were implemented using MATLAB programming language.

## 3. Results

Figure 3 shows the results for segmenting a PCG signal. In the staircase graph of Figure 3, levels 1, 2, 3, and 4 define the S1 sound intervals, the systole intervals, the S2 sound intervals, and the diastole intervals, respectively. As the segmentation strategy we used is a sophisticated strategy proposed and evaluated by Springer et al. (2015) [18], we did not evaluate its performance.

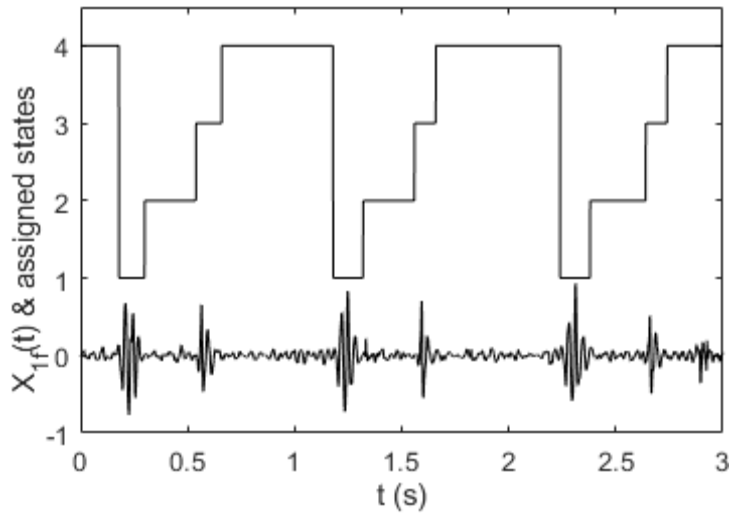

Figure 3: Segmentation of a PCG signal: The staircase graph shows the segmentation results of the plotted PCG signal: level 1 shows S1 intervals, level 2 shows systole intervals, level 3 shows S2 intervals, and level 4 shows diastole intervals.

### 3.1. Results for the single-classifier strategy

The single-classifier was trained and tested using the 52 mean MFCCs extracted from the PCGs, as shown in Figure 2.A. Three types of classifiers were evaluated in this strategy: kNN, SVM, and DT. For the kNN classifier, $k = 3$ obtained better results than 1, 5, and 7 values. For that reason, only the results for the case that $k = 3$ are presented. Also, the polynomial kernel had better results for the SVM classifier than the linear and Gaussian kernels. For that reason, only the results of the SVM with the polynomial kernel are presented. The accuracy, sensitivity, and specificity parameters of the three classifiers (kNN, SVM, and DT) are presented in Table 1. Both the support vector machine ($Acc = 91.95\%, Se = 92.78\%, Sp = 91.14\%$) and the kNN ($Acc = 91.9\%, Se = 91.41\%, Sp = 92.48\%$) has outperformed the decision tree ($Acc = 87.33\%, Se = 86.72\%, Sp = 88.03\%$). There was no statistically significant difference between the accuracy of the SVM and the kNN algorithm. However, the SVM had a higher sensitivity than the kNN, while the kNN had a higher specificity. The 95 % confidence intervals for the sensitivity of the SVM and the kNN were [92.31, 93.25] and [91.11, 91.71], respectively. Also, the 95 % confidence intervals for the sensitivity of the SVM and the kNN were [90.67, 91.61] and [92.03, 92.93], respectively.

Table 1: The results for the single-classifier strategy for the three types of classifiers: kNN ($k = 3$), SVM with a polynomial kernel, and DT.

| Classifier | Acc (%) | Se (%) | Sp (%) | MAcc (%) |
|---|---|---|---|---|
| kNN | 91.9 | 91.41 | 92.48 | 91.95 |
| SVM | 91.95 | 92.78 | 91.14 | 91.96 |
| DT | 87.33 | 86.72 | 88.03 | 87.34 |

### 3.2. Results for the ensemble-classifier strategy

All nine were kNN, SVM, or DT in the ensemble-classifier strategy. Similar to the single-classifier strategy, for the kNN classifier, $k = 3$ obtained better results than 1, 5, and 7 values. For that reason, only the results for the case that $k = 3$ are presented. Also, the polynomial kernel had better results for the SVM classifier than the linear and Gaussian kernels. Therefore, only the results of the SVM with a polynomial kernel are presented. The accuracy, sensitivity, and specificity parameters of the three classifiers (kNN, SVM, and DT) are presented in Table 2. The SVM had the highest accuracy (93.59%) and sensitivity (95.4%) while the kNN had the highest specificity (93.02%). The 95% confidence intervals for the accuracy of the kNN, SVM, and DT were [91.57, 92.11], [93.27, 93.91], [91.89, 92.55], respectively. The 95% confidence intervals for the sensitivity of the kNN, SVM, and DT were [90.63, 91.23], [95.05, 95.75], [93.64, 94.42], respectively. The 95% confidence intervals for the specificity of the kNN, SVM, and DT were [92.62, 93.42], [91.4, 92.22], [90.06, 91], respectively.

Table 2: The results for the ensemble classification strategy for three types of classifiers: kNN ($k = 3$), SVM with a polynomial kernel, and DT.

| Classifier | Acc (%) | Se (%) | Sp (%) | MAcc (%) |
|---|---|---|---|---|
| kNN | 91.84 | 90.93 | 93.02 | 91.98 |
| SVM | 93.59 | 95.40 | 91.81 | 93.61 |
| DT | 92.22 | 94.03 | 90.53 | 92.28 |

### 3.3. Comparison of single-classifier and ensemble-classifier strategies

As shown in Figure 4, when the classifier type was either DT or SVM, the ensemble-classifier strategy achieved a higher accuracy than the single-classifier strategy. In Figure 4, the error bars represent the 95% confidence intervals for the classification accuracy. On average, DT's accuracy improved by 4.89%, while SVM's improved by 1.64%. There was no statistically significant difference between the accuracy of the single- and ensemble-classifier strategies when the kNN was used.

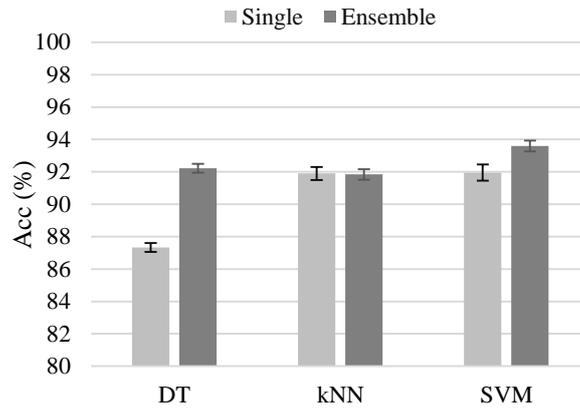

Figure 4. The classification accuracy for the single- (light gray) and the ensemble classifiers (dark gray): the ensemble classifier has outperformed the single classifier for both the decision tree (DT) and the SVM classifier types. The error bars represent the 95% confidence interval for the classification accuracy.

As shown in Figure 5, when the classifier type was either DT or SVM, the ensemble-classifier strategy achieved a higher sensitivity than the single-classifier strategy. In Figure 5, the error bars represent the 95% confidence intervals for the classification sensitivity. On average, DT's sensitivity improved by 7.31%, while SVM's sensitivity improved by 2.62%. There was no statistically significant difference between the sensitivity of the single- and ensemble-classifier strategies when the kNN was used.

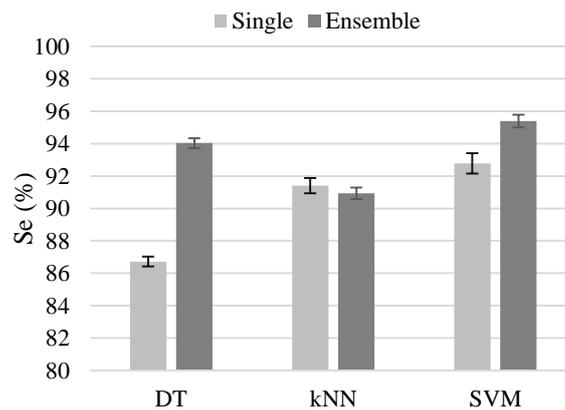

Figure 5. The classification sensitivity for the single- (light gray) and the ensemble classifiers (dark gray): the ensemble classifier has outperformed the single classifier for both the decision tree (DT) and the SVM classifier types. The error bars represent the 95% confidence interval for the classification sensitivity.

As shown in Figure 6, when the classifier type was DT, the ensemble-classifier strategy achieved a higher specificity than the single-classifier strategy. In Figure 6, the error bars represent the 95% confidence intervals for the classification specificity. On average, DT's specificity improved by 2.5%. There was no statistically significant difference between the specificity of the single- and ensemble-classifier strategies when the kNN and SVM were used.

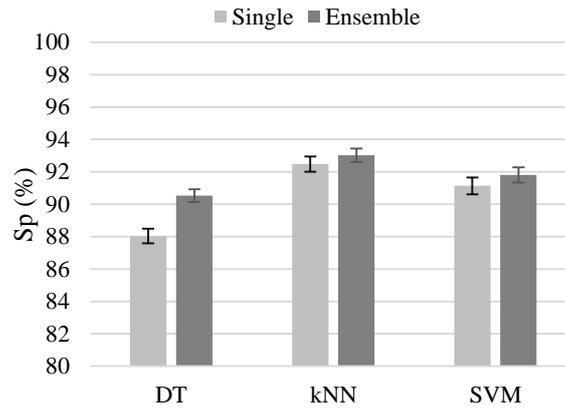

Figure 6. The classification specificity for the single- (light gray) and the ensemble classifiers (dark gray): the ensemble classifier has outperformed the single classifier only for the decision tree (DT) classifier type. The error bars represent the 95% confidence interval for the classification specificity.

## 4. Discussion

The well-known MFCCs have been used in applications such as audio processing [29]. They have also been used for PCG processing [3, 10, 11, 14]. In this study, two classification strategies were designed to investigate the performance of the MFCCs in detecting abnormal PCGs. The single-classifier strategy is based on 52 mean MFCCs extracted from S1, systole, S2, and diastole intervals and determines whether the PCG is normal or abnormal. The ensemble-classifier strategy, however, first determines whether each PCG beat is normal or abnormal using 52 MFCCs of the same PCG beat and then considers the PCG to be normal if it has more normal beats than abnormal ones and vice versa. In the single-classifier strategy, both the SVM ($Acc = 91.95\%$) and the kNN ($Acc = 91.9\%$) obtained higher classification accuracies than the DT ($Acc = 87.33\%$). Still, there was no significant difference between the kNN and SVM.

In the ensemble-classifier strategy, the SVM ($Acc = 93.59\%$) produced a higher classification accuracy than the kNN ($Acc = 91.84\%$) and DT ($Acc = 92.22\%$). Still, there was no significant difference between the kNN and DT.

When comparing the two classification strategies, the accuracy of the SVM and DT improved by 1.64% and 4.89% in the ensemble-classifier strategy, while the accuracy of the kNN did not change significantly. Overall, the accuracy, sensitivity, and specificity of the DT increased considerably from the single-classifier strategy to the ensemble-classifier strategy. Also, the SVM classifier in the ensemble classifier achieved the highest classification accuracy (93.59%). One reason that the ensemble classifier outperformed the single-classifier could be because the averaging of MFCCs across multiple PCG beats in the latter strategy removes the individual beat differences, which are actually very important cues for detecting the abnormal PCGs. Another reason could be that averaging requires a very accurate timing for the PCG segments S1, systole, S2, and diastole with their respective segments in other beats. This accurate timing is almost impossible in practice as phonocardiograms are non-stationary. In a nutshell, the results suggest that the ensemble-classifier strategy is more efficient than the single-classifier strategy.

Table 3 compares the results of the current study with a few similar studies, which have used the PhysioNET CinC 2016 database (section 2.1) and/or MFCCs. The classification accuracies presented in Table 3 were either directly reported or estimated from the data reported in the respective articles. In the approach taken by Khan et al. (2020) [3], MFCCs were estimated from unsegmented signals, leading to a lower accuracy of 80.68%. Unlike that, our method benefits from the segmentation of phonocardiograms into distinct heart sound intervals (S1, systole, S2, and diastole), which likely contributes to our higher accuracies of 91.95% and 93.59% for single and ensemble classifiers, respectively. Though MFCCs were estimated from segmented signals in the method proposed by Milani et al. (2021) [10], S1 and S2 segmentation was based on a simple method of systole and diastole detection, leading to a low accuracy of 83.33%. The concatenation of MFCCs with time domain features increased the accuracy from 83.33% to 93.33%, comparable to the accuracy of 93.59% achieved by our ensemble classifier, increasing the complexity of their model while leaving its specificity (Sp=88.24%) much lower than that of our ensemble-classifier (Se=91.81%).

Features extracted in time, frequency, and time-frequency domains have also been used in applications of PCG processing [3, 8, 12, 13]. Langley and Murray (2017) [8] used Spectral amplitude and wavelet entropy (2 features) to classify unsegmented PCGs, leading to an accuracy of 79.33%, much lower than our single-classifier (Acc=91.95%) and ensemble-classifier strategy (Acc=93.59%), once again confirming that signal segmentation is essential for PCG classification. Unlike the method proposed by Langley and Murray (2017) [8], the method proposed by Khan et al. (2020) [3] classified segmented phonocardiograms using time and time-frequency [3], reaching an accuracy of 91.23%. Similarly, Homsi and Warrick (2017) [13] used time, frequency, statistical, and wavelet features for the classification of segmented phonocardiograms. They reached an accuracy of (Acc=86.58%). Sotaquirá et al. (2018) [12] also used similar features to classify normal/abnormal PCG cycles. They achieved an accuracy of (Acc = 92.6 %) using deep learning. This significantly improves Homsi and Warrick (2017) [13], yet is lower than our ensemble classifier (Acc=93.59%). The superiority of our method (esp. our ensemble classifier), which uses MFCCs, over the methods using time and time-frequency features implies that MFCCs provide a better representation for phonocardiograms than time and time-frequency features.

Table 3. Comparison of the results of this study with similar studies: LPF: low-pass filtering, HPF: high-pass filtering, BPF: band-pass filtering, PIFS: partitioned iterated function systems, MLP: multi-layer Perceptron, DNN: deep neural factor, T: time-domain feature, F: frequency-domain features, T-F: time-frequency features, W: wavelet features, SA: spectral amplitude, WE: wavelet entropy, ST: statistical features

| Study | Preprocessing | Segmentation | Database | Features | Classifier | Results (%) |
|---|---|---|---|---|---|---|
| [3] | BPF | [18] | [26, 27] | T, T-F | ensemble of 100 DTs | Acc: 91.23 |
| [3] | BPF | N/A | [26, 27] | MFCCs | LSTM | Acc: 80.68 |
| [10] | LPF | systole and diastole durations | [26, 27] | MFCCs | ANN | Acc: 83.33 |
| [10] | LPF | systole and diastole durations | [26, 27] | T, MFCCs | ANN | Acc: 93.33 |
| [8] | Zero-mean | N/A | [26, 27] | SA & WE | DT | Acc: 79.33 |
| [13] | BPF | [18] | [26, 27] | T, F, W, ST | ensemble + voting (20) | Acc: 86.58 |
| [12] | BPF | [18] | [26, 27] | T, F, T-F, | DNN | Acc: 92.6 |
| [9] | Savitzky–Golay LPF | N/A | [26, 27] | 1-D CNN | MLP | Acc: 85.65 |
| [9] | Savitzky–Golay LPF | N/A | [26, 27] |  | MLP | Acc: 85.65 |
| [16] | PIFS | N/A | [26, 27] | N/A | CNN | MAcc: 85 |
| [11] | HPF & Amplitude Normalization | Shannon energy, envelope smoothing, peak finding | Private | T, F, T-F, MFCCs | kNN | Acc: 98.78 |
| [14] | BPF | Mel-Scaled Wavelet Transform | Private | T, ST, MFCCs | MLP | Acc: 99.91 |

| Current study | BPF | [18] | [26, 27] | MFCCs | Single-classifier | MAcc: 91.96%<br>Acc: 91.95% |
| --- | --- | --- | --- | --- | --- | --- |
| | | | | | Ensemble-classifier | MAcc: 93.61%<br>Acc: 93.59% |

Some studies investigated the combination of MFCCs with time and time-frequency features to improve classification accuracy. For instance, concatenation of MFCCs with time features increased the classification accuracy from 83.33% to 93.33% in [10]. In [14], concatenating MFCCs with time features and statistical features led to an increased accuracy of 99.91%. Similarly, in [11], concatenating MFCCs with time features, and frequency features, time-frequency features, and wavelet features increased the classification accuracy of tricuspid regurgitation severity using PCGS to 98.78%. Though the PCG databases used in the last two studies differed from the PhysioNET CinC 2016 database, the findings suggest that concatenation of MFCCs with other features might be more effective for PCG applications. However this improved accuracy comes at the cost of increasing the complexity of the systems.

In a different approach taken by Krishnan et al. (2020) [9], a 1-D convolutional network was proposed for feature extraction from unsegmented PCGs, reaching an accuracy of 85.65%, a very low sensitivity of 57.78%, and a specificity of 92.98%. In another experiment, they applied an MLP with 4 hidden layers, increasing the sensitivity from 57.78% to 86.73% while the accuracy remained unchanged. Similarly, Riccio et al. (2023) [16] reached a modified accuracy of 85% using a convolutional neural network. They used Partitioned Iterated Function Systems (PIFS) to generate 2D color images from 1D PCGs. These images were used as input for the CNN. The results of these two studies are much lower than our single-classifier (Acc=91.95%) and ensemble-classifier strategy (Acc=93.59%). This could be because a) they used unsegmented PCGs and/or b) the features extracted by deep neural nets are less efficient than MFCCs.

Figure 7 compares the accuracy, sensitivity, and specificity of the aforementioned methods (cited in Table 3) with our single and ensemble classifiers. Only six of these studies, which a) used the PhysioNET CinC 2016 database [26, 27], and b) reported the accuracy, sensitivity, and specificity directly (or it was possible to estimate them from the reported data), were included in Figure 7. As can be seen in Figure 7, our ensemble classifier has the highest accuracy (Acc=93.59%) and the second highest sensitivity (Se=95.4%). The method proposed in [10] has the highest sensitivity (Se=100%), but it should be emphasized that its specificity (Sp=88.24%) is lower than that of our ensemble classifier (Sp=91.84%). Finally, our ensemble classifier has the third highest specificity (Sp=91.81%). The highest specificities were achieved by [3] (Sp=97.04%) and [12] (Sp=93.8%), but both have lower sensitivities (Se=78.81%, and 91.3%, respectively) than our ensemble classifier (Se=95.40%). Overall, it seems that our ensemble classifier has outperformed other studies.

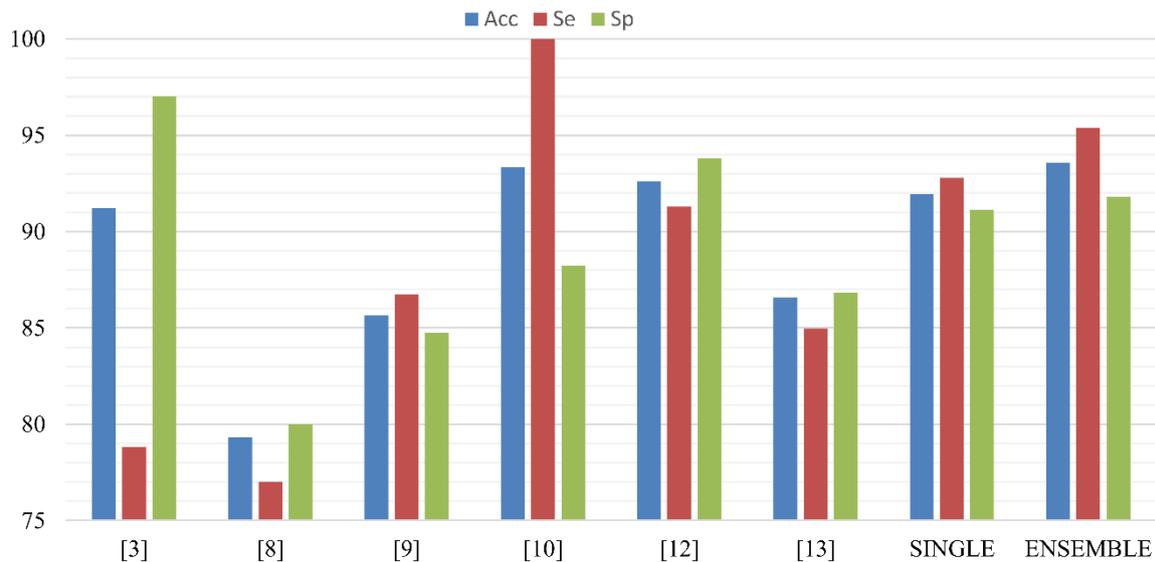

Figure 7. Comparison of the accuracy, sensitivity, and specificity of our ensemble and single classifiers with the some of the methods cited in Table 3.

Though our ensemble-classifier achieved a high accuracy for PCG classification, our method has a number of limitations. First, as the segmentation algorithm we used was state-of-the-art, we did not evaluate the segmentation step. Second, the MFCCs capture only the spectral properties of the heart sounds, while as shown in previous studies [10, 11, 14], temporal features can contribute to the classification performance. Therefore, it is necessary to incorporate the temporal features of the heart sounds into our model in future. Third, since MFCCs were effective for phonocardiogram classification, it should be investigated whether they can effectively be used to develop a supervised segmentation algorithm. If so, the complexity of the proposed strategies will be reduced significantly. Last, although our results confirm that our ensemble classifier is very efficient for binary classification problems to discriminate abnormal phonocardiograms from normal ones, it is still necessary to evaluate it in multi-class classification problems to detect cardiovascular diseases.

## 5. Conclusion

The performance of MFCCs for detecting abnormal PCGs was evaluated using two classification strategies, i.e., a single-classifier strategy and an ensemble-classifier strategy. In the single-classifier strategy, the MFCCs extracted from different PCG beats are first averaged, and the mean MFCCs are then used to classify PCGs. However, in the ensemble-classifier strategy, MFCCs are used by an ensemble of 9 classifiers to classify PCG beats into normal/abnormal beats. In the end, if most beats are classified as normal, the PCG is considered normal; otherwise, it is abnormal. Both strategies were tested on a publicly available database of PCG signals. The results showed that the ensemble-classifier strategy could classify PCGs with a higher accuracy, establishing MFCCs as more effective than other features, including time, time-frequency, and statistical features, evaluated in similar studies.